# Spin Glass-like Phase below ~ 210 K in Magnetoelectric Gallium Ferrite

*Somdutta Mukherjee[1], Ashish Garg[2] and Rajeev Gupta[1,3,\*]*

[1]Department of Physics

[2]Department of Materials Science and Engineering

[3]Materials Science Programme

Indian Institute of Technology Kanpur, Kanpur 208016, India

**Abstract:**

In this letter we show the presence of a spin-glass like phase in single crystals of magnetoelectric gallium ferrite ($GaFeO_3$) below ~210 K via temperature dependent ac and dc magnetization studies. Analysis of frequency dispersion of the susceptibility peak at ~210 K using the critical slowing down model and Vogel-Fulcher law strongly suggests the existence of a classical spin-glass like phase. This classical spin glass behavior of $GaFeO_3$ is understood in terms of an outcome of geometrical frustration arising from the inherent site disorder among the antiferromagnetically coupled Fe ions located at octahedral Ga and Fe sites.

[\*] Corresponding author; FAX: +91-512-2590914; guptaraj@iitk.ac.in

Simultaneous presence of more than one order parameters in single phase magnetoelectric (ME) and/or multiferroic (MF) materials leads to conceptualization of many exciting devices such as multi-state memories, sensors etc.[1,2] Applicability of these materials for such device applications requires a good understanding of the coupling behavior among magnetic, electrical and structural order parameters since such couplings trigger many interesting phenomena in the above materials when studied over temperature,[3] composition[4] and length scales.[5] For instance, materials such as $SrMnO_3$[6] and $EuTiO_3$[7] undergo strain induced ferromagnetic-ferroelectric phase transition below a critical temperature owing to their large magneto-structural coupling. More recently, our work on another magnetoelectric oxide, $GaFeO_3$ (GFO), which is particularly attractive due to tunability of its magnetic transition temperature, has shown the presence of substantial magneto-structural coupling[8] below room temperature (RT). Our results indicated a sudden change in the strength of the interaction across ~ 200 K strongly suggestive of a spin reorientation in this material. Despite scarcity of such observations of spin reorientation particularly associated with spin frustration (very few exceptions e.g. $BiFeO_3$[3] and $YMnO_3$[9]), there is a growing interest in examining spin dynamics in ME and MF materials since it has been proposed that magnetic frustration in some MF systems can result in spiral magnetic ordering inducing ferroelectricity.[10] These aspects make it essential to examine the magnetic behavior of such materials to understand the spin interactions from a practical perspective of material and device design.

GFO simultaneously exhibits ferrimagnetism and piezoelectricity and its magnetic transition can be tuned above RT by manipulating the material's composition i.e. Ga to Fe ratio.[11] Noncentrosymmetric orthorhombic structure ($Pc2_1n$) of GFO has eight formula units per unit-cell with four inequivalent cationic sites: Ga1 ions occupying tetrahedral sites and Ga2, Fe1 and Fe2 ions occupying all the octahedral sites. Ideally GFO is expected to be an antiferromagnet,[12,13] however finite temperature magnetic measurements show it as a ferrimagnet.[12,14] Latter is believed to be an outcome of cationic site disorder due to very small differences in the sizes of $Ga^{3+}$ and $Fe^{3+}$ ions. Detailed structural characterization using X-ray[8,12] and Neutron diffraction[12] rules out any structural phase transition in the ferrimagnetic state. However, our Raman spectroscopic study, as mentioned above, clearly indicates a subtle change in the spin orientation across ~200 K. Occurrence of these two contrasting events is indeed quite intriguing and requires a careful investigation of the spin dynamics. In this letter we report the results of temperature dependent ac susceptibility and dc magnetization measurements to further elucidate hitherto observed spin reorientation near 200 K in GFO. Our results clearly demonstrate the existence of a spin-glass phase in GFO and we show that this is a manifestation of geometrical frustration emanating from cation site disorder.

Single crystals of GFO were flux grown from high purity precursor oxides $Ga_2O_3$ and $Fe_2O_3$ using $Bi_2O_3$ as flux[8,11] yielding dark brown needle shaped crystals with [110]-orientation. Details of structural characterization can be found elsewhere.[8] Further, samples were subjected to temperature dependent ac and dc magnetization measurements using SQUID magnetometer. The measurements were performed under both Field Cooled (FC) and Zero Field Cooled (ZFC) conditions over a temperature range, 2 K to 330 K. In all the measurements, external dc field and the probing ac field were applied along the c-axis of the crystals.

Temperature dependent dc magnetization data of GFO, at fields: 100 Oe and 500 Oe, are shown in Fig. 1(a). On cooling, magnetization increases sharply below $T_c$ ~290 K marking the transition from the paramagnetic (PM) phase to ferrimagnetic (fM) phase, a well established transition. Cooling the sample below $T_c$ results in splitting of FC and ZFC curves. This splitting marks the onset of magnetic irreversibility at a certain temperature, defined as $T_{ir}$ below which bifurcation between FC and ZFC curves starts occurring. Further lowering of temperature leads to the formation of a cusp in ZFC plot at a temperature defined as $T_p$. Bifurcation of ZFC and FC curves is more pronounced at lower field strength where $T_p$ and $T_{ir}$ remained well separated and shifted toward higher temperature. A cusp in the ZFC plot and the distinctive separation of FC and ZFC data at $T_{ir}$ are typical features of spin-glasses.[3,4,9] This is usually explained in terms of spin freezing or change in the spin-ordering leading to a spin-glass like phase formation at low temperatures. However, the splitting of the FC and ZFC curves is not a sufficient evidence to conclude spin-glass nature[15] and it is also often observed in the ferromagnetic regions in many systems, attributed to the pinning of the domain walls.[4]

In materials showing a spin glass behavior, spin interactions lead to a highly irreversible yet metastable state and can be well analyzed by ac magnetization studies.[3,4] There is, in fact, preliminary evidence of magnetic frustration provided by ac susceptibility measurements on polycrystalline GFO.[16] However, a narrow frequency range of 4 kHz – 10 kHz used in the experiments does not conclusively prove spin-glass nature of GFO. This warrants a detailed investigation using ac susceptibility over a reasonably wide temperature and frequency domain. In this context, we first examine the temperature dependence of ac susceptibility in the frequency range of 0.1 to 1000 Hz as shown in Fig. 1 (b) and (c). Upon cooling from 330 K to 2 K, both the $\chi'$ and $\chi''$ display sharp peaks at ~$T_c$. These frequency independent peaks termed as Hopkinson peaks are typical feature in many ferromagnetic materials.[17] With further lowering of temperature, another set of weak and broad peaks (corresponding to spin freezing temperature $T_f$ ~ 210 K) appear in both $\chi'$ and $\chi''$ plots (Fig. 1(b) and (c)). The peak positions shift to higher temperatures with increasing frequency and also their magnitudes depend strongly on frequency. Frequency dispersion of these low temperature susceptibility peaks has also been observed for a variety of other oxides exhibiting spin glass behavior such as $BiFeO_3$,[3] $LuFe_2O_{4+\delta}$[18] and $CaBaFe_4O_7$[19] and as well as in dilute magnetic alloys[20] and has been explained as an indication of presence of short range spin interactions.

In the spin glass state, the slower spin dynamics with decreasing temperature implies that spins take longer time to relax to a relatively stable state i.e. relaxation time increases with decreasing temperature. Dynamic susceptibility measurements can thus be used to distinguish whether GFO is a classical spin glass or a superparamagnet by comparing the initial frequency dependence of $T_f(\omega)$ using the expression ($\Delta p = \Delta T_f/(T_f \Delta \log\omega)$).[3,20] Our measurements show that $T_f$ varies from ~212 K (0.1 Hz) to ~216 K ($10^3$ Hz) in $\chi'$ plot while it varies from 210 K (0.1 Hz) to ~212 K ($10^3$ Hz) in $\chi''$. The calculated peak shift ($\Delta p$) per decade of frequency shift has a value of about 0.005 and 0.003 for $\chi'$ and $\chi''$, respectively which are an order of magnitude lower than those observed for super-paramagnetic systems ($10^{-1}$–$10^{-2}$) while their values match well with those for classical spin glasses,[20,21] suggesting that GFO undergoes spin glass transition below the freezing temperature. Above can further be substantiated by analyzing the frequency dependence of the peaks in $\chi'$ using the conventional critical slowing down model of spin dynamics,[22] i.e.

$$\frac{\tau}{\tau_0} = \left(\frac{T_f(\omega) - T_s}{T_s}\right)^{-z\nu} \qquad (1)$$

where, $T_s$ is spin glass transition temperature determined by the system interactions (at $\omega \to 0$, $T_f(\omega) \to T_s$), $z$ is dynamic critical exponent, $\nu$ is the critical exponent of the correlation length and $\tau_o$ is the shortest relaxation time available to the system.[22] Fig. 2 (a) shows the best fit to the data in the frequency range, 0.1 to 1000 Hz, suggesting that the spin glass behavior in GFO can be well described using critical slowing down model and the fitting yielded following parameters: $T_s = 211 \pm 0.5$ K, $z\nu = 5.5 \pm 1.5$ and $\tau_o \sim 10^{-13}$ s. These values are in good agreement with those reported for well known spin glass and cluster glass systems. The value of $z\nu$ for most classical as well as cluster glass systems lie between 5-10 such as for Ising spin glass $Fe_{0.5}Mn_{0.5}TiO_3$,[23] geometrically frustrated system $LuFe_2O_{4+\delta}$[18] and cluster glass $U_2CuSi_3$.[24] Thus $zv$ alone cannot be used as a decisive parameter to differentiate between the type of spin glasses.

The other criterion to distinguish different kinds of spin glasses is based on the Vogel[25]- Fulcher[26] law relating the relaxation in a spin glass system to the driving frequency and subsequently estimating the activation energy using the expression:

$$\omega = \omega_0 \exp\left[\frac{-E_a}{k_B(T_f - T_0)}\right] \qquad (2)$$

Here, $T_0$ is the Vogel-Fulcher temperature and $k_B$ is the Boltzmann constant. Taking $\omega_0 = 10^{13}$ Hz as calculated earlier, a linear variation of $T_f$ versus $1/\ln(\omega_0/\omega)$ is obtained and the best fit of the experimental data to the eq. 2 (solid line in Fig. 2(b)) yields $T_0 = 202.9$ K and $E_a = 1.66$ $k_BT_s$. This activation energy, $E_a$ is a measure of the energy barrier separating different metastable states accessible to the system. For a canonical spin glass such as Cu-Mn[20] as well as for a geometrically frustrated system $CaBaFe_4O_7$,[19] $\omega_0$ has been reported to be $\sim 10^{13}$ Hz. However, for cluster glass Li doped $CaBaFe_{4-x}Li_xO_7$ ($x = 0.1$ to $x = 0.4$),[19] $U_2CuSi_3$[24] and $La_{0.5}Sr_{0.5}CoO_3$[27] the reported values of $\omega_0$ range between $10^{12}$-$10^{16}$ Hz. The observed scatter in $\omega_0$ for different systems with similar characteristics thus, does not allow us to draw any meaningful conclusions. On the other hand the value of activation energy, $E_a$, appears to exhibit a trend. For instance, the value of $E_a$ is $\sim 2k_BT_s$ for a canonical spin-glass Cu-Mn[20] (Mn $\sim$ 3.3-8 at.%) and $E_a \sim 1.25k_BT_s$ for geometrically frustrated $CaBaFe_4O_7$[19]. In contrast, $E_a$ is quite large for cluster glass systems: $\sim 12\ k_BT_s$ for Li-doped $CaBaFe_{4-x}Li_xO_7$ ($x=0.4$)[19], $\sim 3.1k_BT_s$ for $U_2CuSi_3$[24] and $\sim 7k_BT_s$ for $La_{0.5}Sr_{0.5}CoO_3$.[27] From this, we can infer that GFO is close to being a classical spin-glass with $E_a \sim 1.66\ k_BT_s$ which can further be substantiated by dc field dependence of ac susceptibility data that can differentiate between a classical spin glass from the assemblies of magnetic clusters based on the temperature shift of the

1   susceptibility peak as a function of applied dc field.[28,29] For a classical spin glass, $T_f$ usually

2   shifts toward lower temperatures with increasing applied dc field while for cluster glass, it

3   moves to higher temperature due to the growth of the clusters.[28,29]

4   Fig. 3 depicts the in-phase component of temperature dependent ac susceptibility

5   where an ac magnetic field of magnitude 4 Oe and a driving frequency of 100 Hz were

6   applied with superimposed different dc fields of 0 to 10 kOe. A first glance, the susceptibility

7   ($\chi'$) vs. temperature plot shows the presence of two peaks: a strong peak in the vicinity of 280

8   K corresponding to fM-PM transition and a low temperature peak at ~210 K corresponds to

9   the spin-glass phase. Fig. 3 clearly shows that with increasing dc field, the low temperature

10  peak shifts to low temperatures (from ~ 207 K at zero field to ~ 185 K at 500 Oe)

11  accompanied by decreasing peak amplitude. The peak eventually disappears at ~ 1 T

12  suggesting complete suppression of spin-glass behavior at higher external fields. Such peak

13  shift (of $T_f$) towards lower temperatures with increasing dc field is observed in many classical

14  spin glasses and can be quantitatively described using de Almeida-Thouless (AT) line for an

15  anisotropic Ising spin glass system[30] as expressed by

16  $$H = H_0 \left( 1 - \frac{T_f(H)}{T(0)} \right)^{\frac{3}{2}} \qquad (3)$$

17  where, H is the external applied dc magnetic field, $T_f$ (H) is the field dependent freezing

18  temperature, and $H_o$ determines the boundary of the applied dc magnetic field up to which the

19  spin glass phase can exist. Eq. 3 suggests that a plot of $H^{2/3}$ vs. freezing temperature ($T_f$)

20  would yield the values of $H_o$ and T(0). The plot is shown in inset of Fig. 3 and we obtain $H_o$ =

21  1.2 T and T(0) = 209 K. The above value of field is close to the experimental observations i.e.

22  susceptibility peak disappearing at ~ 1T. A low value of goodness of fit to the AT line in the

23  plot points towards the Ising nature of the spin glass and the spin freezing in GFO is quite similar to that of conventional spin glass systems. Moreover, the ratio of $T_c$ to $T_s \sim 1.4:1$, also supports Ising nature of the present system as postulated by Campbell *et al.*[31]

Usually, a necessary condition to achieve a spin glass state is magnetic frustration which may or may not be associated with disorder. In case of magnetic oxides, there have been a large number of studies on geometrical spin frustration on compounds such as pyrochlores,[19] spinels[18] and garnets[32] where spin glass behavior is an outcome of the formation of a triangular framework of the antiferromagnetically coupled magnetic ions resulting in spin frustration. In order to analyze the origin of the observed spin glass behavior we propose a physical model of geometrically frustrated network of antiferroimagnetically arranged cations in GFO. GFO has an inherent site disorder where some of the Ga sites are occupied by Fe ions and vice-versa. While A-type antiferromagnetic spin ordering ensures that Fe (at Fe1 and Fe2 sites) ions are antiferromagnetically aligned with respect to each other.[13] Site disordering renders some of the Fe ions to occupy Ga sites (primarily Ga2 sites as the occupancy of Fe at Ga1 site is negligible).[12] This leads to the formation of zigzag chain of corner sharing tetrahedral spin network among Fe1, Fe2 and Fe (at Ga2 site) ions leading to spin frustration in the lattice as shown in Fig. 4. Although similar may also happen at Ga1 sites, theoretical calculations and experimental[12,13] data predict it as quite unlikely. It is likely that such spin frustration affects the antiferromagentic spin ordering in GFO leading to a concomitant co-existence of a spin-glass phase.

In summary, using detailed temperature and frequency dependent dc and ac magnetic measurements, we clearly demonstrate the existence of a spin-glass like transition at ~ 210 K in single crystalline gallium ferrite (GFO), in addition to the previously reported paramagnetic to ferrimagnetic transition at ~290 K. We observe that low temperature peak of ac susceptibility shows strong frequency dispersion and analysis of this frequency dispersion using the critical slowing down model and the Vogel-Fulcher law strongly supports the formation of a classical spin-glass like phase. These results are consistent with a recent report[8] on changes in the magneto-structural coupling coefficient of GFO across ~200 K. We argue that the disorder driven geometrically frustrated corner sharing tetrahedral network of Fe ions gives rise to the observed spin glass phase in GFO.

Authors thank Prof. S. Ramakrishnan for permitting the use of SQUID facility at Tata Institute of Fundamental Research, Mumbai and Mr. G.S. Jangam for conducting the measurements. Authors acknowledge the financial support from the Department of Science and Technology and Council for Scientific and Industrial Research, India.

1   **List of figures:**

2   Fig. 1 (color online) (a) ZFC and FC dc magnetization vs. temperature plots at 100 Oe
3   (filled symbols) and 500 Oe (open symbols). $T_{ir}$ represents the temperature at which
4   bifurcation between FC and ZFC curves occurs and the cusp in ZFC plots is marked as $T_p$.
5   Temperature dependent (b) real ($\chi'$) and (c) imaginary ($\chi''$) parts of ac susceptibility data
6   measured at several frequencies. Inset of 1 (b) shows the magnified view of the low
7   temperature peak in $\chi'$ plot.

9   Fig. 2 (color online) (a) Plot of time constant ($\tau$) vs. dynamic spin freezing temperature ($T_f$) of
10  GFO with solid line representing the best fit to equation 1. (b) Plot of $1/\ln(\omega_0/\omega)$ vs. dynamic
11  spin freezing temperature ($T_f$) with solid line representing the best fit to equation 2.

13  Fig. 3 (color online) Plot of temperature dependence of $\chi'$ measured at several applied dc
14  fields H. The figure also shows that at sufficiently large applied dc fields (e.g. H=1T), the low
15  temperature peak corresponding to the spin glass transition disappears completely. Inset
16  shows the plot of dynamic spin freezing temperature ($T_f$) vs. $H^{2/3}$ with the solid line being the
17  best fit to equation 3.

18

19  Fig. 4. (color online) Schematic diagram illustrating geometrical spin frustration in corner
20  sharing tetrahedral network of Fe ions in GFO unit cell. The frustration (marked as '?') arises
21  due to cationic site disorder driving Fe ions to occupy some of the Ga2 sites. For Clarity Ga1
22  ions which have negligible Fe occupancy are removed.

23

24

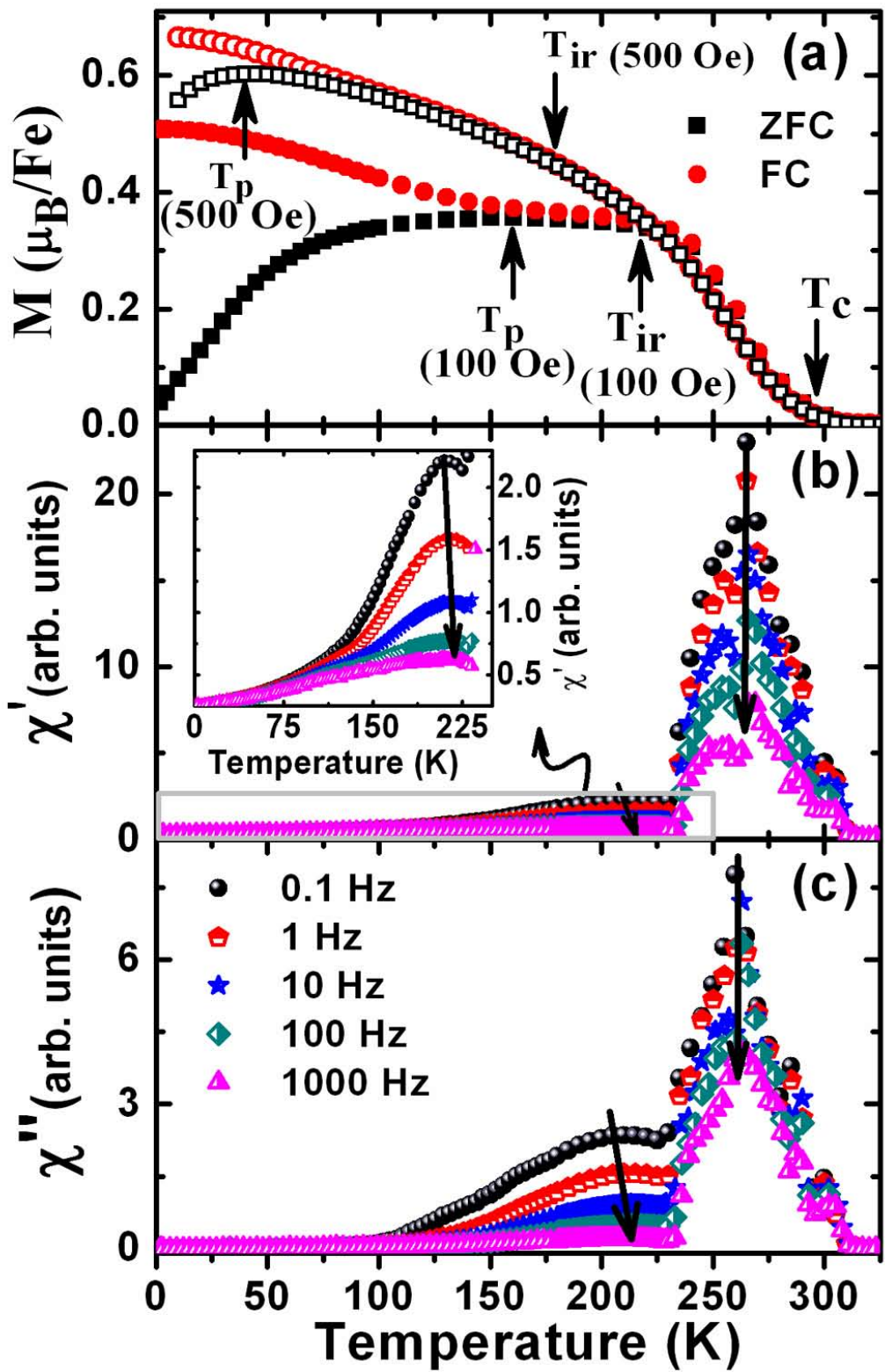

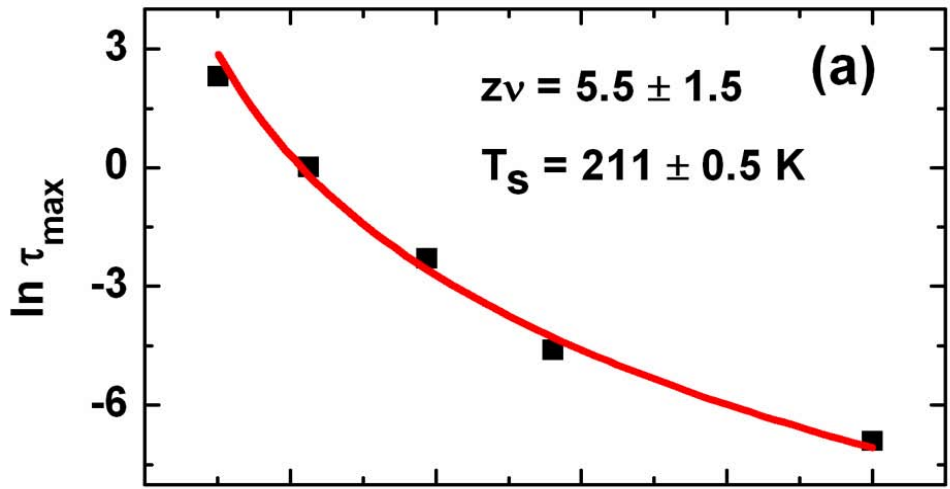

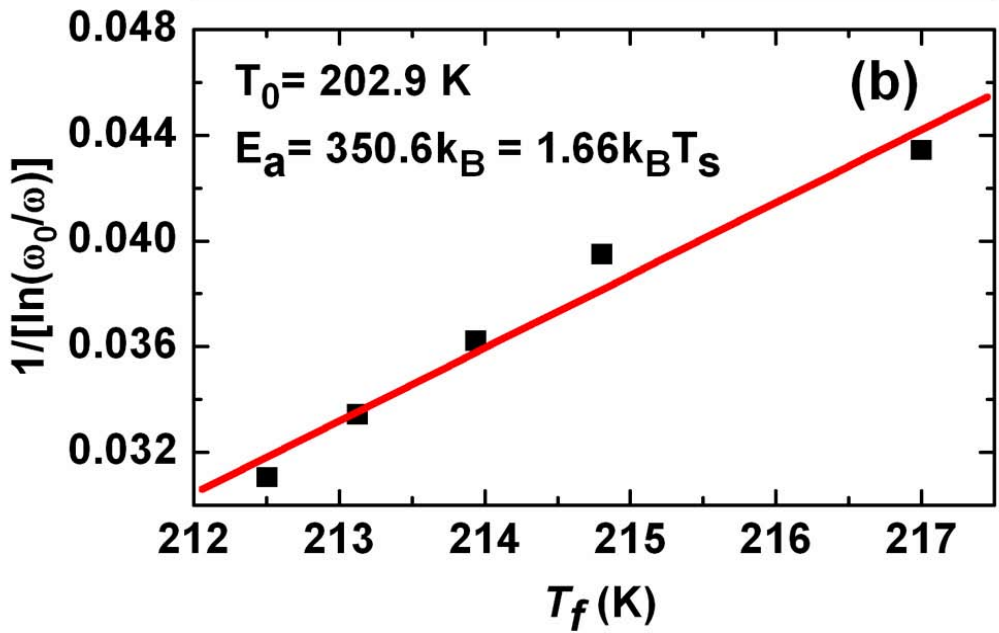

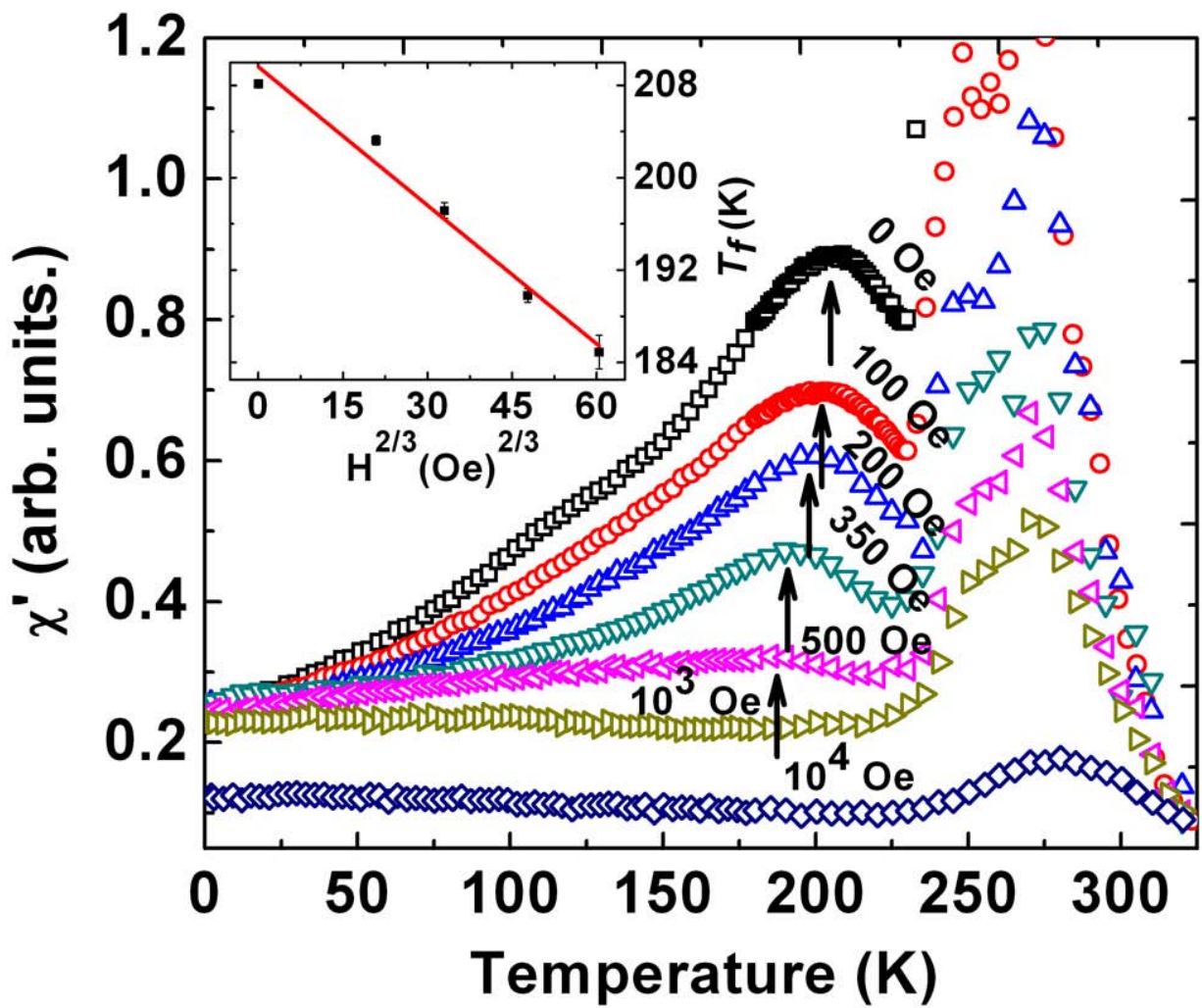

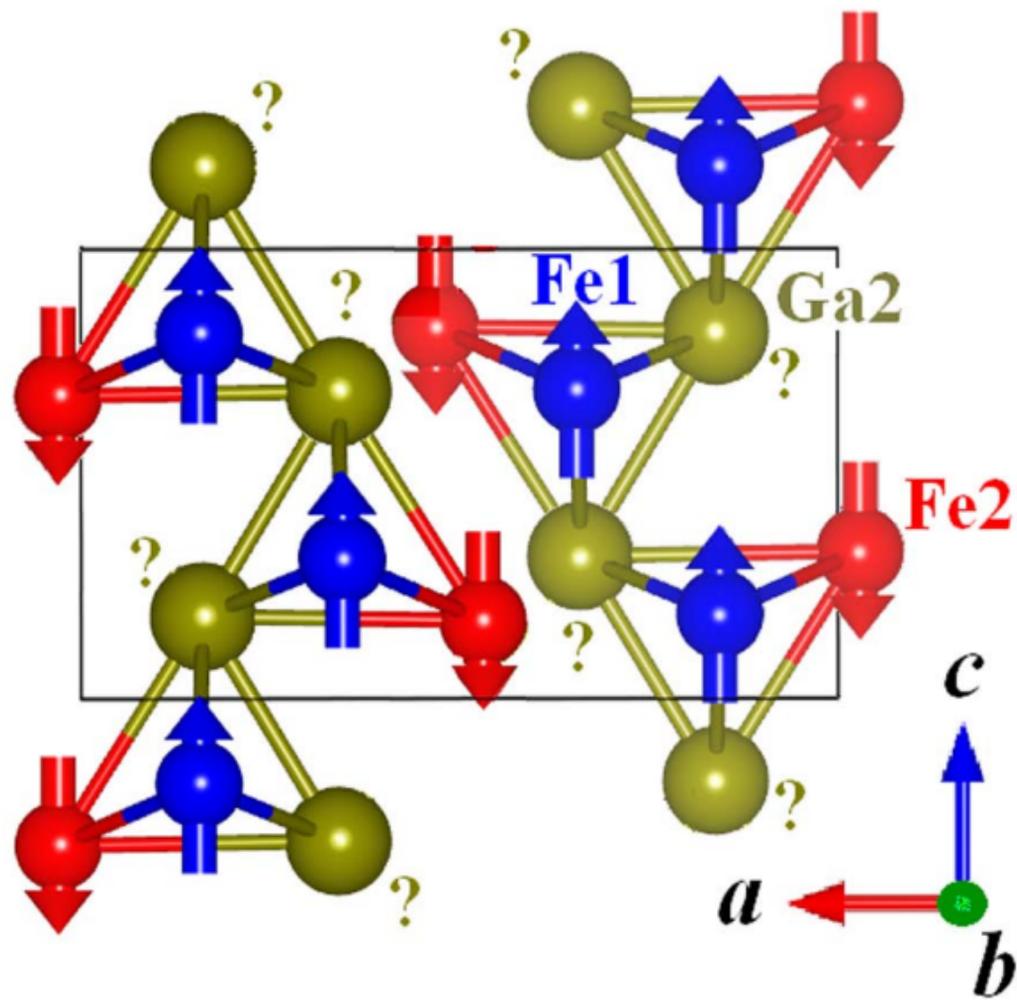